\documentclass[11pt]{article}
\usepackage{cite}
\usepackage{latexsym}
\usepackage{amssymb}
\usepackage{epsf}
\usepackage{amsmath}
\usepackage[hypertex]{hyperref}
\usepackage{graphicx}

\usepackage{slashed}

\newcommand{\tr}{{\rm Tr}}

\begin{document}
\pagestyle{empty}

\begin{flushright}
TU-900
\end{flushright}

\vspace{3cm}

\begin{center}

{\bf\LARGE Making confining strings out of mesons}
\\

\vspace*{1.5cm}
{\large 
Ryuichiro Kitano, Mitsutoshi Nakamura, and Naoto Yokoi
} \\
\vspace*{0.5cm}

{\it Department of Physics, Tohoku University, Sendai 980-8578, Japan}\\
\vspace*{0.5cm}

\end{center}

\vspace*{1.0cm}

\begin{abstract}
{
The light mesons such as $\pi$, $\rho$, $\omega$, $f_0$, and $a_0$ are
possible candidates of magnetic degrees of freedom, if a magnetic dual
picture of QCD exists.
We construct a linear sigma model to describe spontaneous breaking of
the magnetic gauge group, in which there is a stable vortex
configuration of vector and scalar mesons. We numerically examine
whether such a string can be interpreted as the confining string. By
using meson masses and couplings as inputs, we calculate the tension of
the string as well as the strength of the Coulomb force between static
quarks.
They are found to be consistent with those inferred from the quarkonium
spectrum and the Regge trajectories of hadrons.
By using the same Lagrangian, the critical temperature of the QCD phase
transition is estimated, and a non-trivial flavor dependence is
predicted.
We also discuss a possible connection between the Seiberg duality and
the magnetic model we studied.
}
\end{abstract} 

\newpage
\baselineskip=18pt
\setcounter{page}{2}
\pagestyle{plain}
\baselineskip=18pt
\pagestyle{plain}

\setcounter{footnote}{0}

\section{Introduction}

The dual Meissner effect is a plausible explanation of the color
confinement in QCD~\cite{'tHooft:1975pu, Mandelstam:1974vf}.
The condensation of the magnetic monopole, $\langle m \rangle \neq 0$,
makes the QCD vacuum to be in the dual superconducting phase, where
color fluxes sourced by quarks are squeezed into tubes, explaining the
linear potential between quarks.

In the QCD vacuum, there is another interesting phenomenon called the
chiral symmetry breaking. It is believed that the quark-antiquark pair
condenses in the vacuum, and the $SU(N_f)_L$ $\times$ $SU(N_f)_R$
symmetry is spontaneously broken down to a diagonal $SU(N_f)_V$ group.

Since two condensations, $\langle m \rangle \neq 0$ and $\langle \bar q
q \rangle \neq 0$, happen in the same dynamics, these two may be
related. Indeed, lattice simulations of finite temperature QCD suggest
that deconfinement and chiral symmetry restoration happen at similar
temperatures~\cite{Aoki:2009sc, Bazavov:2011nk}.
This leads us to consider a simple unified picture:
\begin{eqnarray}
 \langle \bar q q \rangle \neq 0 
\leftrightarrow \langle \bar m m \rangle \neq 0,
\nonumber 
\end{eqnarray}
where ``$\leftrightarrow$'' represents a non-abelian electric-magnetic
duality.
If the magnetic ``monopoles'' $m$ and $\bar m$ carry flavor quantum
numbers, the condensations $\langle m \rangle = \langle \bar m \rangle
\neq 0$ describe Higgsing of the magnetic gauge group as well as chiral
symmetry breaking. This phenomenon has been observed in supersymmetric
gauge theories~\cite{Seiberg:1994aj, Carlino:2000ff, Konishi:2005qt,
Gorsky:2007ip}.

In this hypothesis, the ``monopole'' condensations give masses to
magnetic gauge bosons and simultaneously provide massless pions as the
Nambu-Goldstone bosons. There is in fact such a structure in the real
hadron world. It has been known that the masses and interactions of the
pions and the vector mesons such as the $\rho$ and the $\omega$ mesons
are well described by a spontaneously broken $U(N_f)$ gauge
theory~\cite{Bando:1984ej}. (See \cite{sakurai} for an earlier
discussion on the description of the vector mesons as gauge fields.)
The vector mesons and the pions are respectively interpreted as the
gauge fields and the uneaten Nambu-Goldstone bosons. Therefore, we are
naturally lead to consider a possibility that the $\rho$ and $\omega$
mesons are actually the magnetic gauge bosons of QCD.
Along this line, it has been demonstrated recently that QCD regularized
into an ${\cal N}=1$ supersymmetric theory has such a magnetic
description via the Seiberg duality~\cite{Seiberg:1994pq}, where the
magnetic Higgs fields $m$ and $\bar m$ are the dual scalar
quarks~\cite{Kitano:2011zk}.

The Higgs model of the $U(N_f)$ gauge theory contains vector and scalar
fields as well as strings as solitonic objects~\cite{Abrikosov:1956sx,
Nielsen:1973cs, Hanany:2003hp, Auzzi:2003fs}. Since the string carries a
magnetic flux in the magnetic picture, it can naturally be identified as
the confining string via the electric-magnetic duality.
We examine in this paper whether such an identification works at the
quantitative level. By using hadron masses and coupling constants as
inputs, one can calculate the string tension and the Coulomb force
between static quarks. We obtain values which are consistent with those
inferred from the quarkonium spectrum and the Regge trajectories in the
hadron spectrum.

We write down a linear sigma model which includes the vector mesons, the
pions and the scalar mesons in the next section.
A vortex configuration in the model is constructed in
Section~\ref{sec:vortex}, and we compare the energy of the
monopole-antimonopole system to the experimentally measured potential
between a quark and an antiquark in Section~\ref{sec:potential}.
The critical temperature of the QCD transition is estimated in
Section~\ref{sec:temp}.
The identification of the vector mesons as magnetic gauge bosons is
motivated by recent discussions in supersymmetric gauge
theories~\cite{Komargodski:2010mc, Kitano:2011zk, Abel:2012un}. (See
also \cite{Harada:1999zj} for an earlier discussion.)
We extend the discussion and propose a new interpretation in
Section~\ref{sec:susy}.

\section{Magnetic linear sigma model}
\label{sec:model}

The magnetic picture of a confining gauge theory is supposed to be a
Higgs model of some gauge theory. We apply this principle in QCD, and
construct a model to describe Higgsing of the magnetic gauge group
as well as chiral symmetry breaking.

\subsection{Lagrangian}
We propose the following Lagrangian to describe the magnetic picture of
QCD. It is a $U(N_f)$ gauge theory, and the Lagrangian possesses the
$U(N_f)_L$ $\times$ $U(N_f)_R$ chiral symmetry. The vacuum expectation
values (VEVs) of the Higgs fields, $H_L$ and $H_R$, break the chiral
symmetry down to the diagonal subgroup, $U(N_f)_V$, providing massless
Nambu-Goldstone bosons identified as pions and $\eta$. The $\eta$ meson
(or $\eta^\prime$ in the three-flavor language) can obtain a mass
through a term which breaks axial $U(1)$ symmetry explicitly such as
$\det (H_L H_R)$ although we ignore it in this paper.
The VEVs of the Higgs fields give masses to $U(N_f)$ gauge bosons. We
identify these massive gauge bosons as the $\rho$ and the $\omega$
mesons\footnote{In the three-flavor language, one should include $K^*
(892)$ and $\phi(1020)$ in the vector mesons.}.
The Lagrangian is given by
\begin{eqnarray}
 {\cal L}
&=& -{1 \over 4} F_{\mu \nu}^{(\omega)} F^{(\omega) \mu \nu}
 -{1 \over 4} F_{\mu \nu}^{(\rho)a} F^{(\rho) \mu \nu a}
\nonumber \\
&&
+ {f_\pi^2 \over 2}
\tr \left[
|D_\mu H_L|^2 + |D_\mu H_R|^2
\right]
\nonumber \\
&& - V(H_L, H_R).
\label{eq:lag}
\end{eqnarray}
The first and the second terms represent the kinetic terms of the $U(1)$
and the $SU(N_f)$ parts of the $U(N_f)$ gauge bosons: $\omega_\mu$ and
$\rho^a_\mu$, respectively.
The Higgs fields $H_L$ and $H_R$ are $N_f \times N_f$ matrices which
transform as
\begin{eqnarray}
 H_L \to g_L H_L g_H^{-1},\ \ \ H_R \to g_H H_R g_R^{-1},
\end{eqnarray}
under the $U(N_f)_L$, the gauged $U(N_f)$, and the $U(N_f)_R$ group
elements, $g_L$, $g_H$, and $g_R$, respectively. The covariant
derivatives are, therefore, given by
\begin{eqnarray}
 D_\mu H_L = \partial_\mu H_L + i g_2 H_L \rho_\mu^a T^a
+ i g_1 Q \omega_\mu H_L,
\end{eqnarray}
\begin{eqnarray}
 D_\mu H_R = \partial_\mu H_R - i g_2 \rho_\mu^a T^a H_R
- i g_1 Q \omega_\mu H_R.
\end{eqnarray}
Here, we normalized the $SU(N_f)$ generators in the fundamental
representation, $T^a$, and the $U(1)$ charge, $Q$, such that
\begin{eqnarray}
 \tr \left( T^a T^b \right) = {1 \over 2} \delta^{ab},
\end{eqnarray}
and
\begin{eqnarray}
 Q = \sqrt{{1 \over 2N_f} }.
\end{eqnarray}

The most general potential terms consistent with the symmetries are
given by
\begin{eqnarray}
 V(H_L, H_R) &=& f_\pi^4 \Bigg [
{\lambda_0 - \lambda_A \over 8 N_f}
\left(
\tr(H_L H_L^\dagger) + \tr(H_R^\dagger H_R) - 2 N_f
\right)^2
\nonumber \\
&&
+  {\lambda_A \over 8}
\left \{ 
\tr \left[
(H_L^\dagger H_L + H_R H_R^\dagger)^2
\right]
- 4 \left(
\tr(H_L H_L^\dagger) + \tr(H_R^\dagger H_R)
\right)
\right \}
\nonumber \\
&&
+ {\lambda^\prime - \lambda^{\prime \prime} \over 8 N_f}
\left(
\tr (H_L H_L^\dagger ) - \tr (H_R^\dagger H_R)
\right)^2
\nonumber \\
&&
+ {\lambda^{\prime \prime} \over 8}
\tr \left[
(H_L^\dagger H_L - H_R H_R^\dagger )^2
\right] \Bigg ],
\end{eqnarray}
where we assumed the parity invariance under $H_L \leftrightarrow H_R$.
This potential stabilizes $H_L$ and $H_R$ at
\begin{eqnarray}
 \langle H_L \rangle = 
\langle H_R \rangle = {\bf 1}.
\end{eqnarray}
At the vacuum, $4 N_f^2$ degrees of freedom in $H_L$ and $H_R$ break up
to $N_f^2$ massless Nambu-Goldstone bosons, $N_f^2$ longitudinal modes
of the gauge bosons, $N_f^2$ massive scalar particles, and $N_f^2$
massive pseudoscalar particles. The decay constant of the
Nambu-Goldstone particles is given by $f_\pi$ at tree level. The gauge
group is completely broken and the unbroken global symmetry is vectorial
$U(N_f)_V$.

The physical modes at the vacuum can be classified by the
representations of $U(N_f)_V$, the spin and the parity.
The masses of the physical modes are given by
\begin{eqnarray}
 \mbox{singlet Nambu-Goldstone boson $(\eta)$:}\ \ \ m_\eta = 0,
\end{eqnarray}
\begin{eqnarray}
 \mbox{adjoint Nambu-Goldstone boson $(\pi)$:}\ \ \ m_\pi = 0,
\end{eqnarray}
\begin{eqnarray}
 \mbox{singlet vector $(\omega)$:}\ \ \ m_\omega^2 = g_1^2 f_\pi^2,
\end{eqnarray}
\begin{eqnarray}
 \mbox{adjoint vector $(\rho)$:}\ \ \ m_\rho^2 = g_2^2 f_\pi^2,
\end{eqnarray}
\begin{eqnarray}
 \mbox{singlet scalar ($f_0$):}\ \ \ m_S^2 = 2 \lambda_0 f_\pi^2,
\label{eq:scalar-mass}
\end{eqnarray}
\begin{eqnarray}
 \mbox{adjoint scalar ($a_0$):}\ \ \ m_A^2 = 2 \lambda_A f_\pi^2,
\label{eq:ad-scalar-mass}
\end{eqnarray}
\begin{eqnarray}
 \mbox{singlet pseudoscalar:}\ \ \ m_{PS}^2 
= 2 \lambda^\prime f_\pi^2,
\end{eqnarray}
\begin{eqnarray}
 \mbox{adjoint pseudoscalar:}\ \ \ 
m_{PA}^2 = 2 \lambda^{\prime \prime} f_\pi^2,
\end{eqnarray}
at tree level. 
Terms with $\lambda^\prime$ and $\lambda^{\prime \prime}$ are not very
important in the following discussion\footnote{ The pseudoscalar
particles are, in fact, $CP$ even, and thus they are exotic states which
are absent in the hadron spectrum. One should take large $\lambda^\prime$ and
$\lambda^{\prime \prime}$ to make the exotic states heavy so that the
model can be a low-energy effective theory of QCD. We thank M.~Harada,
V.A.~Miransky, and K.~Yamawaki for discussion on this point.}.

Hereafter, we take 
\begin{eqnarray}
 g_1 = g_2 \equiv g,
\end{eqnarray}
as the $\rho$ and $\omega$ mesons have similar masses.

\subsection{Vector mesons and pions}

When we integrate out the massive scalar and pseudoscalar mesons, the
model reduces to a non-linear sigma model of
Ref.~\cite{Bando:1984ej}. The matching at tree level gives $a=1$, where
$a$ is a parameter in the low-energy Lagrangian:
\begin{eqnarray}
 {\cal L} \ni {(1-a) f_\pi^2 \over 4} {\tr} \left[
| \partial_\mu (U_L U_R) |^2
\right].
\end{eqnarray}
The unitary matrices $U_L$ and $U_R$ are fields to describe the
Nambu-Goldstone modes including the ones eaten by the gauge bosons. The
transformation properties of $U_L$ and $U_R$ under the gauge and flavor
groups are the same as $H_L$ and $H_R$, respectively.
 From the low energy data, the preferred value of $a$ is estimated to be
$a \sim 2$ with an error of 15\%~\cite{Harada:2003jx}.
Although there is a factor of two difference from the prediction, this
discrepancy can be explained by including quantum corrections and/or
higher dimensional operators. As discussed in Ref.~\cite{Harada:2003jx},
the quantum correction makes the Lagrangian parameter $a(\Lambda)$
approaches to unity when we take $\Lambda$ to be large, such as $a(
\Lambda ) \simeq 1.33 \pm 0.28$ for $\Lambda = 4\pi f_\pi \sim 1$~GeV.
Moreover, the quantum corrections from the scalar loops give positive
contributions to the gauge boson masses, that further reduces the
$a(\Lambda)$ parameter. Therefore, one can think of the Lagrangian in
Eq.~\eqref{eq:lag} as the one defined at a high energy scale such as the
mass scale of the scalar mesons.

However, the large quantum corrections result in predictions which
depend on the choice of input physical quantities when we work at tree
level, although the differences should be canceled after including
quantum corrections.
In this case, one should choose a set of physical quantities which gives
small enough coupling constants so that the use of the perturbative
expansion is valid and the tree-level results are reliable.

The Lagrangian has four parameters relevant for the discussion: $g$,
$f_\pi$, $\lambda_0$, and $\lambda_A$. The $\lambda_0$ and $\lambda_A$
parameters can be obtained from the scalar masses as we discuss
later. The gauge coupling constant $g$ and the $f_\pi$ parameter can be
estimated from two of physical quantities. The well-measured physical
quantities which can be used as input parameters
are~\cite{Harada:2003jx}:
\begin{eqnarray}
 g_{\rho} = (340~{\rm MeV})^2, \ \ \ 
 g_{\rho \pi \pi} = 6.0,\ \ \ 
 F_\pi = 92~{\rm MeV}, \ \ \ 
 m_\rho = 770~{\rm MeV},
\label{eq:inputs}
\end{eqnarray}
where $g_\rho$ and $g_{\rho \pi \pi}$ are the decay constant and the
coupling to two pions of the $\rho$ meson measured by $\rho \to e^+ e^-$
and $\rho \to \pi \pi$ decays, respectively, and $F_\pi$ is the decay
constant of the pion.
The relations to the Lagrangian parameters at tree level are given by
\begin{eqnarray}
 g_\rho = g f_\pi^2 ,\ \ \ 
 g_{\rho \pi \pi} = {g \over 2} , \ \ \ 
 F_\pi = f_\pi , \ \ \ 
 m_\rho = g f_\pi .
\end{eqnarray}
Among them, the pair to give the smallest gauge coupling is $g_\rho$ and
$m_\rho$ such as
\begin{eqnarray}
 g = {m_\rho^2 \over g_\rho} = 5.0, \ \ \ 
 f_\pi = {g_\rho \over m_\rho} = 150~{\rm MeV}.
\label{eq:parameters}
\end{eqnarray}
The value $g = 5.0$ means that the loop expansion parameter, $g^2 N_f /
(4\pi)^2$, is of order 30\% whereas other choices of input quantities
give $90 - 210$\% for $N_f = 2$. Therefore, the choice above is unique
to make a quantitative prediction.
Indeed, the values in Eq.~\eqref{eq:parameters} are close to the ones
evaluated at one-loop level. In Ref.~\cite{Harada:2003jx}, the
parameters at a scale $\Lambda \sim 1$~GeV is obtained to be $g
(\Lambda) \sim 3.3-4.2$, $f_\pi (\Lambda) \sim 130-150$~MeV, and
$a(\Lambda) \sim 1.0 - 1.5$, which reproduce all the physical quantities
in Eq.~\eqref{eq:inputs}.
We use the values of $g$ and $f_\pi$ in Eq.~\eqref{eq:parameters} in the
following discussion. 
However, we should bear in mind that there are theoretical uncertainties
at the level of a factor of two in the results obtained at the classical
level.

\subsection{Scalar mesons}

In the hadron spectrum, there are light scalar mesons, such as $\sigma$,
$\kappa$, $f_0(980)$ and $a_0(980)$, which have not been understood as
$q \bar q$ states in the quark model since they are anomalously light.
We propose to identify them as the Higgs bosons in this linear sigma
model.
We do not consider heavier scalar mesons as candidates since otherwise
the formulas in Eqs.~\eqref{eq:scalar-mass} and
\eqref{eq:ad-scalar-mass} indicate that the coupling constants are large
and the perturbation theory would not be applicable.

By taking the masses of $f_0(980)$ and $a_0(980)$ as input
quantities\footnote{Since $\sigma$ and $\kappa$ are quite broad
resonances, we do not use their masses as inputs.}, {\it i.e.,}
\begin{eqnarray}
 m_S = m_A = 980~{\rm MeV},
\end{eqnarray}
Eqs.~\eqref{eq:scalar-mass} and \eqref{eq:ad-scalar-mass} give the
coupling constants $\lambda_0$ and $\lambda_A$ as
\begin{eqnarray}
 \sqrt {\lambda_0} = \sqrt {\lambda_A} = 4.6,
\label{eq:lambda}
\end{eqnarray}
at tree level, where $f_\pi$ in Eq.~\eqref{eq:parameters} is used.  We
use these values of coupling constants for later calculations. 

\section{Vortex strings}
\label{sec:vortex}

Since the model has a spontaneously broken gauged $U(1)$ factor, there
is a vortex string as a classical field configuration.
The string carries a quantized magnetic flux.
Below we construct a solution with a unit flux, which will be identified
as the confining string.

There have been similar approaches to the confinement in QCD. The
Ginzburg-Landau models (the magnetic Higgs models) are constructed from
phenomenological approaches~\cite{Nielsen:1973cs, Nambu:1974zg,
Jevicki:1974sk} or based on the QCD Lagrangian~\cite{Suzuki:1988yq,
Maedan:1989ju} through the abelian projection~\cite{'tHooft:1981ht}, and
the stable vortex configurations are identified as the confining string.
In supersymmetric theories, there have been numbers of discussion on the
vortex configurations~\cite{Strassler:1997ny, Hanany:2003hp,
Auzzi:2003fs, Eto:2006dx, Eto:2007hf, Shifman:2007kd, Shifman:2011ka,
Hanaki:2011vb}. In particular, the non-abelian
string~\cite{Hanany:2003hp, Auzzi:2003fs}, which we discuss shortly, has
been extensively studied as a candidate of the confining string.

Our model combines Higgsing of the magnetic gauge group and chiral
symmetry breaking. As discussed in the previous Section, the model
parameters are fixed by physical quantities such as masses and couplings
of hadrons. Therefore, the properties of the strings such as the string
tension can be evaluated quantitatively. Below, we explicitly construct
a classical field configuration of the vortex string.

\subsection{Non-abelian vortex solutions}
In this model, there are string configurations called the non-abelian
vortices which carry the minimal magnetic flux.
By defining the following gauge field,
\begin{eqnarray}
 A_\mu^{ij} = {\sqrt 2} \left(
Q \omega_\mu \delta_{ij} + T^a_{ij} \rho^a_\mu
\right),
\end{eqnarray}
there is a vortex configuration made of, {\it e.g.,} the $i=j=1$
component rather than the overall $U(1)$ gauge field $\omega_\mu$. Compared
to the string solution made of $\omega_\mu$, this non-abelian string carries
only $1/N_f$ of the magnetic flux and thus it is stable.

In constructing the vortex configurations, we follow the formalism and
numerical methods of Ref.~\cite{Ball:1987cf}, where the potential
between a monopole and an anti-monopole is evaluated numerically in the
abelian-Higgs model. Classical field configurations are constructed
by numerically solving field equations while imposing the gauge field to
behave as the Dirac monopoles~\cite{Dirac:1948um} as approaching to
their locations.

We consider a non-abelian vortex solution, where the magnetic flux is
sourced by a Dirac-monopole and a Dirac-antimonopole configurations of
the $A_\mu^{ij}$ gauge field with $i=j=1$, representing non-abelian
monopole configurations.
These monopole and anti-monopole are not present as physical states in
the model of Eq.~\eqref{eq:lag}, and we introduce them as field
configurations with an infinite energy, {\it i.e.,} static
quarks\footnote{In $U(N)$ gauge theories with Higgs fields in the
adjoint representation, there are monopoles as solitonic objects which
are identified as junctions of vortices~\cite{Shifman:2004dr,
Hanany:2004ea} rather than the endpoints. The monopoles we are
considering should not be confused with such configurations. }. The
object we construct here, therefore, corresponds to a bound state of
heavy quarks such as the charmonium and the bottomonium. In order to
describe light mesons, the light quarks should be present somewhere in
the whole framework. We discuss a possible framework in
Section~\ref{sec:susy}.

In the cylindrical coordinate, $(\rho, \varphi, z)$, where the monopole
and the antimonopole located on the $z$-axis at $z= \pm R/2$, we denote
$(A_D)_\mu^{ij}$ as the configuration to describe the
monopole-antimonopole system. They are given by
\begin{eqnarray}
 (A_D)_0^{ij} = 0,
\end{eqnarray}
\begin{eqnarray}
  {\boldsymbol A}_D^{ij} = 0, \ \ \ \mbox{except for}\ \ i = j = 1,
\end{eqnarray}
and
\begin{eqnarray}
 {\boldsymbol A}_D^{11} 
= a_D \hat{\boldsymbol \varphi}
= - {N_{\rm flux} \over \sqrt 2 g}{1 \over \rho}
\left[
{z - R/2 \over [
\rho^2 + (z-R/2)^2
]^{1/2}}
-
{z + R/2 \over [
\rho^2 + (z+R/2)^2
]^{1/2}}
\right] 
\hat{\boldsymbol \varphi}.
\end{eqnarray}
The number of the flux, $N_{\rm flux}$, is quantized as $N_{\rm flux}
\in {\mathbb Z}$ by the Dirac quantization
condition~\cite{Dirac:1948um}. Equivalently, the magnetic charge of the
monopole is quantized as
\begin{eqnarray}
 q_m = {4 \pi N_{\rm flux} \over \sqrt 2 g}.
\label{eq:mag-charge}
\end{eqnarray}
The gauge field is well-defined everywhere except for the interval $-R/2
\leq z \leq R/2$ on the $z$-axis. The Dirac quantization condition
ensures that the interval is covered in a different gauge.
For constructing a vortex configuration, the following ansatz are
taken:
\begin{eqnarray}
 A_\mu^{ij} = A_\mu^i \delta^{ij},\ \ \ 
 A_\mu^{i} = (A_D)_\mu^{ii} + a_\mu^{i},
\end{eqnarray}
\begin{eqnarray}
 a_0^i = 0,\ \ \ 
 {\boldsymbol a^i} = a^i (\rho, z) 
\hat{\boldsymbol \varphi},
\end{eqnarray}
\begin{eqnarray}
 (H_L)_{ij} = (H_R)_{ij} = \phi_i (\rho,z) \delta_{ij},\ \ \ 
 \phi_i = \phi_i^{*}.
\end{eqnarray}

With the ansatz, the Lagrangian is reduced to
\begin{eqnarray}
 {\cal L} 
&=& -{1 \over 4} \sum_i F^{i\mu \nu} F^i_{\mu \nu}
\nonumber \\
&&
+ f_\pi^2 \sum_i (\partial_\mu \phi_i )^2
+ {f_\pi^2 \over 2} g^2 \sum_i \phi_i^{2} (A_\mu^i)^{2}
\nonumber \\
&&
- {\lambda_0 \over 2 N_f} f_\pi^4
\left(
\sum_i \phi_i^{2} - N_f
\right)^2
\nonumber \\
&&
- {\lambda_A \over 2 N_f} f_\pi^4
\left(
N_f \sum_i \phi_i^{4} - \left(\sum_i \phi_i^{2}\right)^2
\right)
,
\end{eqnarray}
and the field equations are obtained as
\begin{eqnarray}
 \nabla^2 \phi_i
- {g^2 \over 2} ({a^i + a_D \delta^{i1}})^2 \phi_i
= {\lambda_0 \over 2 N_f} \left(
\sum_j \phi_j^2 - N_f
\right) \phi_i
+ {\lambda_A \over 2 N_f} \left(
N_f \phi_i^2 - \sum_j \phi_j^2
\right) \phi_i,
\label{eq:eom}
\end{eqnarray}
\begin{eqnarray}
 \left(
\nabla^2 - {1 \over \rho^2} \right)
a^i
= {g^2 \over 2} (a^i + a_D \delta^{i1}) \phi_i^2,
\end{eqnarray}
where we take the unit of 
\begin{eqnarray}
 \sqrt 2 f_\pi = 1.
\label{eq:unit}
\end{eqnarray}
For $i\neq 1$, $a^i=0$ is the solution.

The potential energy between the monopole and the anti-monopole is given
by
\begin{eqnarray}
 V(R) &=& 
- {2 \pi N_{\rm flux}^2 \over g^2  R}
\nonumber \\
&&
+
\int d^3 x
\left[
- {g^2 \over 4} \phi_1^2 (a^1 + a_D) a^1
- {\lambda_0 \over 8 N_f} \left(
\left(
\sum_i \phi_i^2 \right)^2 - N_f^2
\right)
\right.
\nonumber \\
&&
\hspace{2cm}
\left.
- {\lambda_A \over 8 N_f} \left(
N_f \sum_i \phi_i^4 - \left( \sum_i \phi_i^2 \right)^2
\right)
\right].
\end{eqnarray}
The first term comes from the magnetic Coulomb potential, $V_{\rm
Coulomb} = - q_{\rm mag}^2 / 4 \pi R$. The second term is the
contribution from the non-trivial field configurations, and gives the
linear potential between a monopole and an antimonopole for a large
$R$. The self-energies of the Dirac monopoles are subtracted, and thus
this expression provides a finite quantity.

For $\lambda_0 = \lambda_A$, which is the case as in
Eq.~\eqref{eq:lambda}, the problem simplifies to the case of the abelian
string. The field equations gives
\begin{eqnarray}
 \phi_i = 1,\ \ \ \mbox{for }  i\neq 1,
\end{eqnarray}
as solutions and the equations for $\phi_1$ and $a^1$ becomes
\begin{eqnarray}
 \nabla^2 \phi_1
- {g^2 \over 2} (a^1 + a_D)^2 \phi_1
= {\lambda_0 \over 2} \left(
\phi_1^2 - 1
\right) \phi_1,
\label{eq:scalar}
\end{eqnarray}
\begin{eqnarray}
 \left(
\nabla^2 - {1 \over \rho^2} \right)
a^1
= {g^2 \over 2} (a^1 + a_D) \phi_1^2.
\label{eq:vector}
\end{eqnarray}
The potential energy is in this case given by
\begin{eqnarray}
  V(R) &=& 
- {2 \pi N_{\rm flux}^2 \over g^2  R}
+
\int d^3 x
\left[
- {g^2 \over 4} \phi_1^2 (a^1 + a_D ) a^1
- {\lambda_0 \over 8} \left(
\phi_1^4 - 1
\right)
\right].
\label{eq:potepote}
\end{eqnarray}
The $N_f$ dependence disappears from the potential
energy.

\subsection{Numerical results}

We numerically solve Eqs.~\eqref{eq:scalar} and \eqref{eq:vector} by
following the procedure explained in Ref.~\cite{Ball:1987cf}. The
partial differential equations are solved by using the Gauss-Seidel
method. The obtained field configurations are used to evaluate the
potential energy in Eq.~\eqref{eq:potepote}.

\begin{figure}[t]
\begin{center}
 \includegraphics[width=10cm]{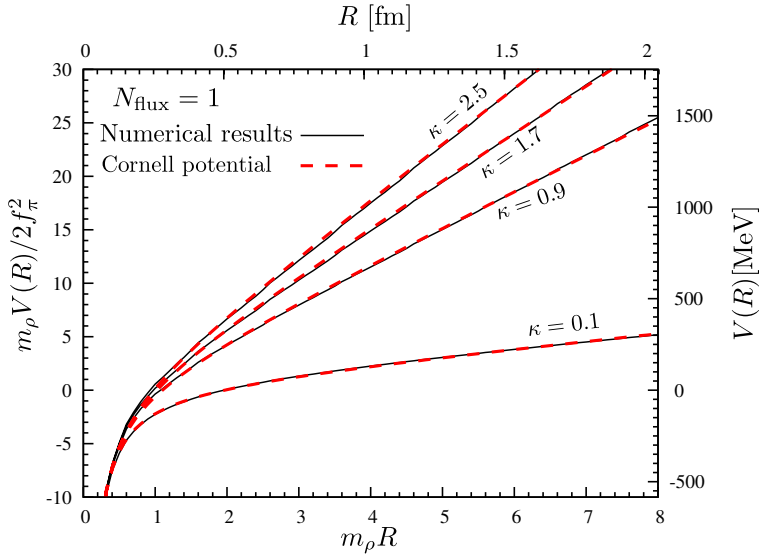}
\caption{Potential energy of the monopole-antimonopole system for 
$\kappa = 0.1$, 0.9, 1.7 and 2.5. The fittings with the 
Cornell potential are superimposed (dashed lines).}
\label{fig:energy}
\end{center}
\end{figure}

\begin{figure}[h]
\begin{center}
 \includegraphics[width=9cm]{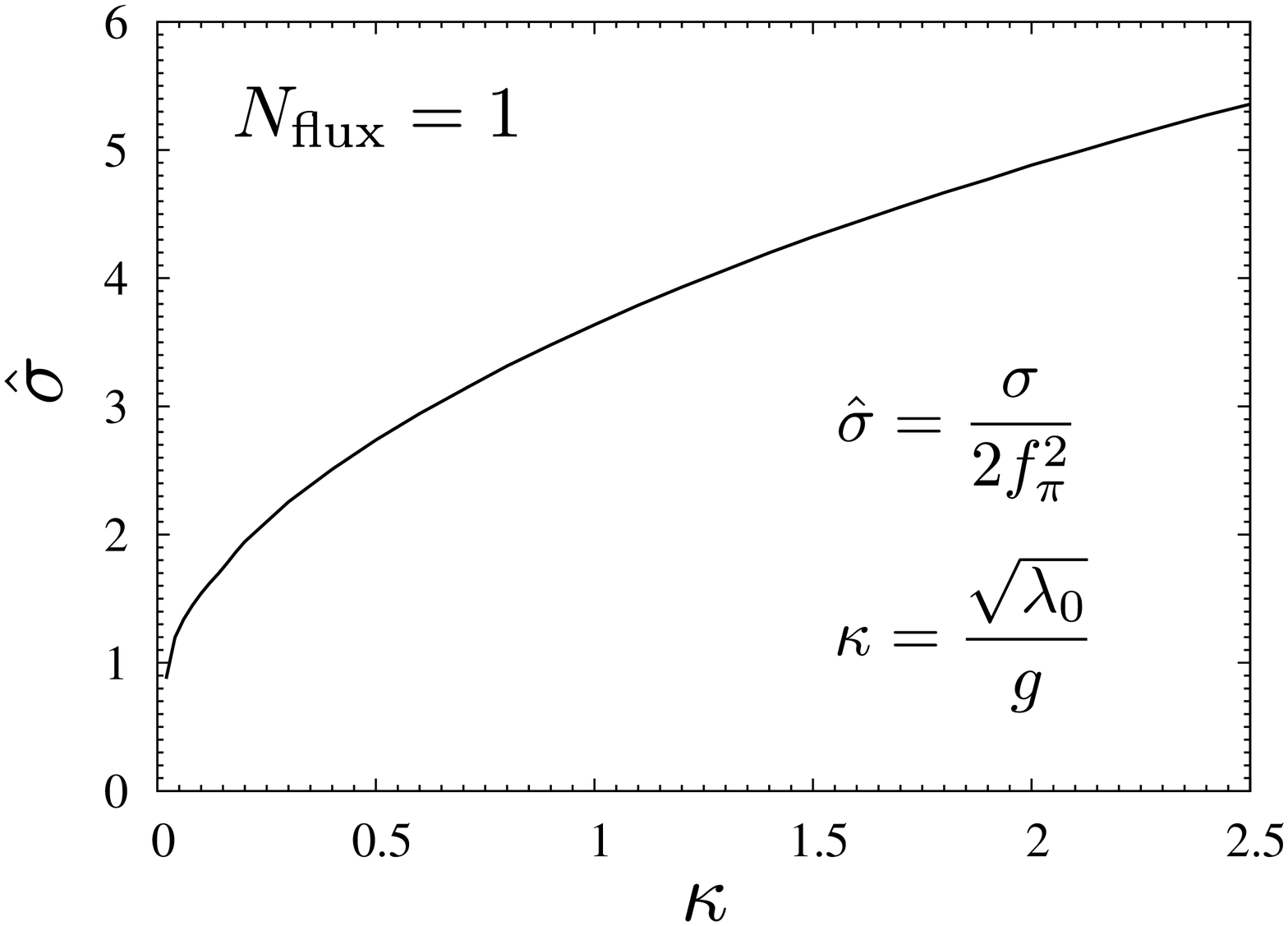}
\caption{The string tension $\hat \sigma$ in the unit of $2 f_\pi^2$
as a function of the Ginzburg-Landau parameter $\kappa$.}
\label{fig:tension}
\end{center}
\end{figure}

In the unit of Eq.~\eqref{eq:unit}, the potential energy $V(R)$ times
the gauge boson mass $m_\rho$ can be obtained as a function of $m_\rho
R$. In this normalization, we have a single parameter $\kappa$ defined
by
\begin{eqnarray}
 \kappa = {m_S \over \sqrt{2} m_\rho} = {\sqrt {\lambda_0} \over g}
= {\sqrt {\lambda_A} \over g}.
\label{eq:kappa}
\end{eqnarray}
This corresponds to the Ginzburg-Landau parameter of superconductors.
The numerical results are shown in Fig.~\ref{fig:energy}, where the
potential energies for $N_{\rm flux} = 1$ are drawn with four choices
of parameters, $\kappa = 0.1$, 0.9, 1.7, and 2.5.
We see a linear potential in a large $R$ region.
By fitting the slope of the linear regime, one can extract the string
tension $\hat \sigma$ in the unit of Eq.~(\ref{eq:unit}). We show in
Fig.~\ref{fig:tension} the tension $\hat \sigma$ as a function of
$\kappa$. These results are all consistent with Ref.~\cite{Ball:1987cf},
except that the unit of the flux is different due to the non-abelian
feature of the vortex.
For $\kappa = 1/\sqrt 2$, the field equations reduce to a set of first
order differential equations whose solutions are known as the BPS
state. In this case, the tension is simply given by $\hat \sigma = \pi$,
which we have confirmed with an accuracy of $0.1-0.2$ percent.


\section{Comparison to QCD data}
\label{sec:potential}

Now we compare the numerical results with data from experimental
measurements. We identify the non-abelian Dirac monopoles with the
minimal magnetic charge, $N_{\rm flux}=1$, as static quarks, since
otherwise the string with $N_{\rm flux} = 1$ is stable and such a stable
string is absent in QCD.
The potential between a quark and an antiquark with a distance $R$ can
be parametrized by the following form:
\begin{eqnarray}
 V(R) = - {A \over R} + \sigma R.
\label{eq:cornell}
\end{eqnarray}
This potential, called the Cornell potential, well fits the quarkonium
spectrum with parameters:
\begin{eqnarray}
 A \sim 0.25-0.5 , \ \ \ 
 \sqrt \sigma \sim 430~{\rm MeV}.
\label{eq:coulomb}
\end{eqnarray}
A similar value of the string tension $\sigma$ is obtained from the
Regge trajectories of the hadron spectrum. The lattice simulations also
reproduce the shape of the potential with $A \sim
0.25-0.4$~\cite{Takahashi:2002bw, Aoki:2002uc, Aoki:2008tq} and $\sqrt
\sigma / m_\rho \sim 0.50-0.55$~\cite{Karsch:2000kv}.
In perturbative QCD, at tree level, the Coulomb part $V \sim -A/R$ is
obtained from the one-gluon exchange between quarks.
At a higher loop level, the shape of the potential approaches to the
form in Eq.~\eqref{eq:cornell}~\cite{Sumino:2003yp}.
Computations at three-loop level have been performed recently in
Refs.~\cite{Anzai:2009tm, Smirnov:2009fh}, and it is reported that the
result is in good agreement with lattice simulations up to a distance
scale $R \lesssim 0.25$~fm~\cite{Anzai:2009tm}.
See, for example, Ref.~\cite{Bali:2000gf} for a review of the static QCD
potential.

The Cornell potential also well fits the numerically obtained potential
in the previous section. We superimpose the fittings with the Cornell
potential in Fig.~\ref{fig:energy} as dashed lines.

\subsection{Coulomb potential}
In the electric picture, {\it i.e.,} in QCD, the Coulomb part $V \sim
-A/R$ is obtained with
\begin{eqnarray}
 A = {N_c^2 - 1 \over 2 N_c} {g_s^2 \over 4 \pi},
\end{eqnarray}
where the strong gauge coupling $g_s$ depends on $R$ through
renormalization.

By duality, in the magnetic picture, the Coulomb term is accounted by a
magnetic Coulomb force between monopoles.
By using the magnetic charge in Eq.~\eqref{eq:mag-charge} with $N_{\rm
flux} = 1$, the coefficient is given by
\begin{eqnarray}
 A = {q_m^2 \over 4 \pi} = {2 \pi \over g^2}.
\end{eqnarray}
This corresponds to the first term in Eq.~\eqref{eq:potepote}.
 Using the value of $g$ in Eq.~\eqref{eq:parameters}, we obtain
\begin{eqnarray}
 A = 0.25.
\label{eq:A}
\end{eqnarray}
The value is consistent with Eq.~\eqref{eq:coulomb}.  This is already an
interesting non-trivial test of the hypothesis that the vector mesons
are the magnetic gauge fields.

Note here that the Coulomb term in Eq.~\eqref{eq:potepote} arises from
a solution of the classical field equations with boundary conditions
given by the Dirac monopoles. Although the vacuum is in a Higgs phase,
the Coulomb force dominates when the distance $R$ is small compared to
the inverse of the gauge boson mass.
In the world-sheet theory of the string, it has been known that the
Coulomb force can be reproduced as the L{\" u}scher term which stems
from the boundary conditions of the string world
sheet~\cite{Luscher:1980ac}. Interestingly, the L{\" u}scher term gives
$A=\pi/12 \sim 0.26$ which is pretty close to the above estimation.

\subsection{Linear potential}

As we have seen already, the linear potential is obtained as in
Fig.~\ref{fig:energy}. The normalized string tension $\hat \sigma$ is
shown in Fig.~\ref{fig:tension} as a function of $\kappa$.
 From Eqs.~\eqref{eq:parameters}, \eqref{eq:lambda} and
 \eqref{eq:kappa}, the $\kappa$ parameter is given by
\begin{eqnarray}
 \kappa = 0.90.
\end{eqnarray}
With this value, we obtain from Fig.~\ref{fig:tension},
\begin{eqnarray}
 \hat \sigma = 3.5.
\label{eq:sigmahat}
\end{eqnarray}
By using $f_\pi$ in Eq.~\eqref{eq:parameters} to recover the mass
dimension, we obtain
\begin{eqnarray}
 \sqrt \sigma = 400~{\rm MeV}.
\end{eqnarray}
This is close to $\sqrt \sigma$ in Eq.~\eqref{eq:coulomb}.
The prediction is not very sensitive to $\kappa$. For example, $\kappa =
0.6 - 1.2$ gives $\sqrt \sigma = 360 - 420$~MeV.
Although we expect a large theoretical uncertainty from quantum
corrections, it is interesting to note that the estimated string tension
is in the right ballpark. The hypothesis that the $\rho$ and $\omega$
mesons as magnetic gauge bosons and light scalar mesons as the Higgs
bosons is found to be consistent with the experimental data.

It is important to notice that there is no dependence on $N_f$ in the
field equations~\eqref{eq:scalar}, \eqref{eq:vector} or in the
expression of the QCD potential~\eqref{eq:potepote}. It is essential to
have this property that the string is non-abelian.
The dimensionless quantity $\sqrt \sigma / m_\rho$ is, in this case,
predicted to be $N_f$ independent, which is consistent with the results
from the lattice QCD~\cite{Karsch:2000kv}.

\section{QCD phase transition}
\label{sec:temp}

At a finite temperature, QCD phase transition takes place. The lattice
simulations support that deconfinement and chiral symmetry restoration
happen at similar temperatures. The chiral transition temperature has
been computed in lattice simulations, and found to be $T_c \sim
150-160$~MeV~\cite{Aoki:2009sc, Bazavov:2011nk} for physical quark
masses.

A simple estimate of the transition temperature is possible in the
magnetic model in Eq.~\eqref{eq:lag}. 
The deconfinement and the restoration of the chiral symmetry both
correspond to the phase transition to the vacuum with $H_L = H_R = 0$,
which is stabilized by thermal masses at a finite temperature.
When we define the transition temperature $T_c$ to be the one at which
the Higgs fields become non-tachyonic at the origin, the temperature is
obtained to be~\cite{Dolan:1973qd}
\begin{eqnarray}
 T_c = \sqrt{8 \over \eta N_f} f_\pi,
\label{eq:tc}
\end{eqnarray}
where the factor $\eta$ is a dimensionless quantity given by
\begin{eqnarray}
 \eta = 1 + {2 m_\rho^2 \over m_S^2} 
+ {2 m_{PS}^2 + m_S^2 \over 3 m_S^2},
\label{eq:eta}
\end{eqnarray} 
at the lowest level of perturbation.  Each term in the $\eta$
parameter represents the contribution to the thermal masses of the Higgs
fields from different particles. The first term, the unity, is the
contribution from the scalar mesons. One should add up all the particles
which obtain masses from the VEVs of $H_L$ and $H_R$.
The estimation of $\eta$ is quite non-trivial since there are particles
which we did not consider, such as nucleons, and also the summation
should be weighted by the abundance in the thermal bath, which may be
affected by their large thermal masses, {\it i.e.,} there may be large
higher order corrections.

By putting $f_\pi$ in Eq.~\eqref{eq:parameters}, we obtain
\begin{eqnarray}
 T_c = \left \{
\begin{array}{ll}
 170~{\rm MeV} \times \left(
\displaystyle{ \eta \over 3 }
\right)^{-1/2}, & (N_f = 2), \\
140~{\rm MeV} \times \left(
\displaystyle{ \eta \over 3 }
\right)^{-1/2}, & (N_f = 3). \\
\end{array}
\right.
\end{eqnarray}
The value $\eta \sim 3$ seems to give temperatures consistent with ones
from lattice simulations. It is interesting that $\eta \sim 3$ is
obtained from Eq.~\eqref{eq:eta} when we take $m_{PS}$ around the cut-off scale, $\Lambda \sim 1$~GeV.

The formula in Eq.~\eqref{eq:tc} predicts that the transition
temperature is inversely proportional to $\sqrt{ N_f }$. This is
numerically consistent with the flavor dependence of $T_c$ studied in
Ref.~\cite{Karsch:2000kv} for two and three flavors in the chiral limit.
There, $T_c$ is obtained to be $173 \pm 8$~MeV and $154 \pm 8$~MeV for
two and three flavors, respectively.
A simulation with a larger number of $N_f$ should be able to test this
prediction.

\section{Non-supersymmetric duality from the Seiberg duality}
\label{sec:susy}

The assumption in the whole framework is the electric-magnetic duality
between the $SU(N_c)$ gauge theory with $N_f$ massless quarks and
$U(N_f)$ gauge theory with bosonic Higgs fields.
The replacement of $N_c$ in the gauge group with $N_f$ is familiar in
supersymmetric gauge theories. For example, the Seiberg duality in the
${\cal N } = 1$ supersymmetric theories replaces $SU(N_c)$ gauge group
by $SU(N_f - N_c)$ in the magnetic picture. We explain here a possible
connection between the Lagrangian in Eq.~\eqref{eq:lag} and the Seiberg
duality, which is discussed in Ref.~\cite{Kitano:2011zk}. We extend the
discussion of Ref.~\cite{Kitano:2011zk} regarding the vortex string and
interpretations of constituent quarks.

It is obvious that the non-supersymmetric QCD can be obtained from
supersymmetric QCD's by adding masses to superpartners and send them to
infinity. What is non-trivial is if a vacuum in the theory with small
masses of superpartners is continuously connected to the
non-supersymmetric theory when we send the masses to large values. Such
a continuous path may or may not exist depending on the space of
parameters defined by a supersymmetric theory to start with.
Recently, it is found in Ref.~\cite{Kitano:2011zk} that there is an
explicit model which reduces to QCD in a limit of parameters and has a
vacuum with the same structure as the low energy QCD in a region of
parameters where the Seiberg duality can be used.
By hoping that the region extends to the QCD limit, one can study
non-perturbative features of QCD, such as strings, at the classical
level in the dual picture.

The proposed mother theory is ${\cal N}=1$ supersymmetric QCD with $N_c$
colors and $N_f + N_c$ flavors.
By giving supersymmetric masses to the extra $N_c$ flavors and soft
supersymmetry breaking masses for gauginos and scalar quarks, one
obtains non-supersymmetric QCD with $N_c$ colors and $N_f$ flavors. The
global symmetries and quantum numbers are listed in
Table~\ref{tab:ele-K}, where $SU(N_c)$ is the gauge group.
The $U(1)_{B^\prime}$ symmetry is absent in the actual QCD, and will be
spontaneously broken in the vacuum we discuss later. In order to avoid
the appearance of the unwanted Nambu-Goldstone mode associated with this
breaking, we gauge $U(1)_{B^\prime}$. The $SU(N_c)_V$ group is also an
artificially enhanced symmetry, and thus we gauge it.
Since the added gauge fields only interact with extra flavors, the limit
of large mass parameters still gives the non-supersymmetric QCD we
wanted.

\renewcommand{\arraystretch}{1.3}
\begin{table}[t]
\begin{center}
\small
 \begin{tabular}[t]{cccccccc}
& $SU(N_c)$ & $SU(N_f)_L$ & $SU(N_f)_R$ & $U(1)_B$ 
& $SU(N_c)_V$ &  $U(1)_{B^\prime}$ & $U(1)_R$ 
\\ \hline \hline
 $Q$ & $N_c$ & $N_f$ & $1$ & 1 & 1 & 0 &
$(N_f - N_c)/N_f$ \\
 $\overline Q$ & $\overline {N_c}$& 1 & $\overline {N_f}$ & $-1$ 
& 1 & 0 & $(N_f - N_c)/N_f$ \\ \hline
 $Q^\prime$ & $N_c$ & 1 & $1$ & 0 & $\overline{N_c}$ & 1 &
1 \\
 $\overline Q^\prime$ & $\overline {N_c}$& 1 & 1 & 0 
& ${N_c}$ & $-1$ & 1 \\ 
 \end{tabular}
\end{center}
\caption{Quantum numbers in the electric picture.}
\label{tab:ele-K}
\end{table}
\renewcommand{\arraystretch}{1}

The magnetic picture of the mother theory is an $SU(N_f)$ gauge theory
with $N_f + N_c$ flavors and meson fields. The particle content and the
quantum numbers are listed in Table~\ref{tab:mag-K}. It was found in
Ref.~\cite{Kitano:2011zk} that there can be a stable vacuum outside the
moduli space by the help of the soft supersymmetry breaking terms. The
vacuum is at $\langle q \rangle = \langle \bar q \rangle \neq 0$, where
$SU(N_f) \times SU(N_f)_L \times SU(N_f)_R$ is spontaneously broken down
to a single vectorial $SU(N_f)_V$ symmetry, that is the isospin
symmetry. The symmetry breaking provides massless pions and
simultaneously gives masses to the $SU(N_f) \times U(1)_{B^\prime}$
gauge fields. Those massive gauge fields can be identified as the vector
mesons, $\rho$ and $\omega$.

\renewcommand{\arraystretch}{1.3}
\begin{table}[t]
\begin{center}
\hspace*{-.8cm}
\small
 \begin{tabular}[t]{ccccccccc}

& $SU(N_f)$ & $SU(N_f)_L$ & $SU(N_f)_R$ & $U(1)_B$ 
& $SU(N_c)_V$  & $U(1)_{B^\prime}$ & $U(1)_R$ 
\\ \hline \hline
 $q$ & $N_f$ & $\overline{N_f}$ & 1 & 0 
& 1 & $N_c/N_f$ &  $N_c/N_f$ \\ 
 $\overline q$ & $\overline {N_f}$ & 1 & $N_f$ & 0 
& 1 & $-N_c/N_f$ &  $N_c/N_f$ \\ 
 $\Phi$ & 1 & $N_f$ & $\overline{N_f}$ & 0 
& 1 & 0 &  $2 (N_f - N_c)/N_f$ \\ \hline
 $q^\prime$ & $N_f$ & 1 & 1 & 1 
& ${{N_c}}$ & $-1 +  N_c/N_f$ &  0 \\ 
 $\overline q^\prime$ & $\overline {N_f}$ & 1 & 1 & $-1$ 
& $\overline{N_c}$ & $1 - N_c/N_f$ &  0 \\ 
 $Y$ & 1 & 1 & $1$ & 0 & 1 + Adj. & 0 &
2 \\
 $Z$  & 1 & 1 & $\overline {N_f}$ & $-1$ 
& $\overline{N_c}$ & 1 & $(2 N_f - N_c)/N_f$ \\ 
 $\overline Z$ & 1 & $N_f$ & 1 & 1 & ${N_c}$ & $-1$ &
$(2 N_f - N_c)/N_f$ \\

 \end{tabular}
\end{center}
\caption{Quantum numbers in the magnetic picture.}
\label{tab:mag-K}
\end{table}
\renewcommand{\arraystretch}{1}

Although the deformation with massive $N_c$ flavors provides us with a
QCD-like vacuum, there are several unsatisfactory features as noted in
Ref.~\cite{Kitano:2011zk}.
Here we discuss those issues and consider a possible interpretation.
In the above discussion, it sounds somewhat strange that the
$U(1)_{B^\prime}$ gauge field is identified as the $\omega$ meson which
is in the same nonet as the $\rho$ meson, whereas the $U(1)_{B^\prime}$
seems to have a completely different origin from the $SU(N_f)$ magnetic
gauge group.
Second, in the particle content in Table~\ref{tab:mag-K}, there are
fields which have $U(1)_B$ charges $\pm 1$, {\it i.e.,} ``quarks.''
These degrees of freedom do not match the picture of confinement since
they look like free quarks.
Finally, there is a vortex string associated with the spontaneous
breaking of $U(1)_{B^\prime}$, which we would like to identify as the
QCD string. However, since the stability of the string is ensured by
topology, it is stable even in the presence of the massless quarks. The
real QCD string should be unstable since a pair creation of the quarks
can break the string.

A possible interpretation is emerged from the consideration of the
origin of $U(1)_{B^\prime}$ in the magnetic picture.
As one can notice from the quantum numbers, $U(1)_{B^\prime}$ in the
electric and magnetic pictures look different.
In particular, the gauged global symmetry in the electric picture is
$U(N_c) \simeq (SU(N_c) \times U(1))/{\mathbb Z}_{N_c} $ whereas one cannot
find a $U(N_c)$ gauge group in the magnetic picture.
This leads us to consider a possibility that there is an additional
$U(1)$ factor as a part of the magnetic gauge group. The actual magnetic
gauge group is $U(N_f)$, and it is broken by a VEV of a field with the
quantum number of $Q^{\prime N_c} q^{N_f}$ so that $U(1)_{B^\prime}$ in
the magnetic picture is an admixture of two $U(1)$'s. Namely, the
duality of the gauge group goes through an intermediate step:
\begin{eqnarray}
 SU(N_c) \times U(N_c) \ \ \mbox{(electric)}
&\to& U(N_f) \times U(N_c)  \ \ \mbox{(magnetic)}
\nonumber \\
&\to& SU(N_f) \times
  SU(N_c)_V \times U(1)_{B^\prime}  \ \ \mbox{(magnetic)}.
\label{eq:int-step}
\end{eqnarray}
Under this assumption, when we send the gauge coupling of $U(1) (\subset
U(N_c))$ in the electric picture to be a large value, the gauge boson of
the $U(1)_{B^\prime}$ factor in the magnetic picture is mostly the one
from the $U(N_f)$ magnetic gauge group. The identification of the
$\omega$ meson becomes reasonable since the origin is now the same as
the $\rho$ meson.

Since $U(1)_{B^\prime}$ is spontaneously broken by $\langle q \rangle =
\langle \bar q \rangle \neq 0$, there is a stable vortex string which
can be explicitly constructed as a classical field configuration in the
magnetic picture. 
The duality steps~\eqref{eq:int-step} imply that there is another string
in the magnetic picture: one associated with $U(N_f)$ and another with
$U(N_c)$.
However, if we go back to the electric picture, there is only a single
$U(1)$ factor in $U(N_c)$, which can only give a single kind of string.
This sounds like a mismatch of two descriptions.

We propose here that the $U(N_f)$ string, made of $q$, $\bar q$, $\rho$,
and $\omega$, is in fact unstable since the ``quarks'' can attach to the
endpoints, and thus that is the one which should be identified as the
QCD string. The $U(N_c)$ string is stable, but should decouple in the
QCD limit.
As mentioned already, there are ``quarks'' in the magnetic picture,
$q^\prime$, $\bar q^\prime$, $Z$ and $\bar Z$. They are natural
candidates of the ``quarks'' which attach to the $U(N_f)$ string.
In turn, if they are the degrees of freedom at the string endpoints, a
linear potential prevents them to be in the one-particle
states. Therefore, the ``quarks'' disappear from the spectrum.
This interpretation seems to give resolutions to all the unsatisfactory
features raised before: the nature of $\omega$, free quarks, and the
stable string.

For this interpretation to be possible, $q^\prime$, $\bar q^\prime$, $Z$
and $\bar Z$ should carry magnetic charges of $U(N_f)$ in addition to
the quantum numbers listed in Table~\ref{tab:mag-K}. Since we assume the
electric-magnetic duality between the $SU(N_c)$ and the $U(N_f)$ gauge
groups, it is equivalent to say that $q^\prime$, $\bar q^\prime$, $Z$
and $\bar Z$ should be colored under $SU(N_c)$, {\it i.e.,} $Z$ and
$\bar Z$ are the quarks (the non-abelian monopoles in the magnetic
picture) and $q^\prime$ and $\bar q^\prime$ are non-abelian dyons. It is
interesting to notice that they indeed have $N_c$ degrees of freedom.

In the $SU(N_f) \times U(1)_{B^\prime}$ magnetic gauge group, there is a
$U(1)$ factor which rotates a particular component of $q_I$ and $\bar
q_I$, where $I$ is the index of the $SU(N_f)$ gauge group. The vortex
string associated with such a $U(1)$ factor is called the non-abelian
string and the one with the minimal magnetic flux is stable. Therefore,
the ``quarks'' should attach to this string. When we take $q_1$ is the
one which rotates under the $U(1)$ factor and normalize the charge of it
as unity, the charges of other charged fields are listed in the left
column of Table~\ref{tab:dyon}.
By assuming that $q$ and $\bar q$ have no magnetic charges, the
Dirac-Schwinger-Zwanziger condition~\cite{Schwinger:1966nj,
Zwanziger:1969by} allows the magnetic charges listed in the right column
of Table~\ref{tab:dyon} as the minimal magnetic charges divided by $(2
\pi / e)$ with $e$ being the gauge coupling constant. Interestingly,
they agree with the ``color charge'' of $SU(N_c)_V$ up to a
normalization, which may be indicating that a part of $SU(N_c)_V$ in the
magnetic picture descends from the electric gauge group, $SU(N_c)$.
For dynamical fields with both electric and magnetic quantum numbers, we
loose the standard Lagrangian description of the model. However, since
the sector of $q$, $\bar q$ (and $\Phi$) is all singlet under
$SU(N_c)_V$ and is decoupled from the colored sector, there can be a
Lagrangian to describe it, and we assume that is the model in
Eq.~\eqref{eq:lag}.

\renewcommand{\arraystretch}{1.3}
\begin{table}[t]
\begin{center}
\hspace*{-.8cm}
\small
 \begin{tabular}[t]{ccc}

& (electric charges)$/e$ & (magnetic charges)$/(2 \pi / e)$ 
\\ \hline \hline
 $q_1$ & 1 & 0 \\ 
 $\bar q_1$ & $-1$ & 0 \\ 
\hline
 $q^\prime_1$ & $1 - 1/N_c$ & $1$ \\ 
 $\bar q^\prime_1$ & $-1 + 1/N_c$ & $-1$ \\ 
 $q^\prime_{I \neq 1}$ & $ - 1/N_c$ & $1$ \\ 
 $\bar q^\prime_{I \neq 1}$ & $ 1/N_c$ & $-1$ \\ 
 $Z$ & $ 1/N_c $ & $-1$ \\ 
 $\bar Z$ & $- 1/N_c$ & $1$ \\

 \end{tabular}
\end{center}
\caption{Electric and magnetic charges under a $U(1)$ factor in $SU(N_f)
\times U(1)_{B^\prime}$.}  \label{tab:dyon}
\end{table}
\renewcommand{\arraystretch}{1}

It is amusing to see that many ingredients to describe the hadron world
are present in this model, such as the vector mesons, the pions, the
light scalar mesons, the QCD string, and the constituent quarks. This is
somewhat surprising since the Seiberg duality is supposed to describe
only massless degrees of freedom. The non-trivial success of the model
may be indicating that the addition of $N_c$ massive quarks is a right
direction to fully connect the electric and magnetic pictures of ${\cal
N}=1$ supersymmetric QCD.

\section*{Acknowledgments}

We would like to thank Yukinari Sumino for valuable information on the
static QCD potential in perturbative QCD. RK would also like to thank
Yutaka Ookouchi for useful discussion. RK is supported in part by the
Grant-in-Aid for Scientific Research 23740165 of JSPS. NM is supported
by the GCOE program ``Weaving Science Web beyond Particle-Matter
Hierarchy.''

\end{document}